\documentclass{elsart}

\usepackage{epsfig}
\usepackage{subfigure}

\begin{document}

\begin{frontmatter}


\title{The duration of recessions follows an exponential not a power law}
\author{Ian Wright}
\ead{wright@ikuni.com}
\ead[url]{ianusa.home.mindspring.com}
\address{iKuni Inc.,\\ 3400 Hillview Avenue, Building 5, Palo Alto, CA 94304, USA\\Fax: +1 650 320 9827\\Phone: +1 650 320 5355}

\title{}


\author{}

\address{}

\begin{abstract}
Ormerod and Mounfield \cite{ormerod01} analyse GDP data of 17 leading
capitalist economies from 1870 to 1994 and conclude that the frequency
of the duration of recessions is consistent with a power-law. But in
fact the data is consistent with an exponential (Boltzmann-Gibbs) law.
\end{abstract}

\begin{keyword}
Econophysics \sep Power-laws \sep Exponential-laws


\PACS 64.60.Lx \sep 64.60.Ht  
\end{keyword}
\end{frontmatter}


Ormerod and Mounfield \cite{ormerod01} analyse GDP data of 17 leading
capitalist economies from 1870 to 1994 and measure the frequency of
duration of recessions. The duration of a recession is defined as the
number of years in which real GDP growth (measured as a percentage
change) is less than zero. Table 1 reproduces their data.

\begin{table}[h]
\caption{Duration of recessions}
\begin{tabular}{l l l l l l l l}
\hline
Duration (yr) & 1 & 2 & 3 & 4 & 5 & 6 & 7 \\
\hline
Frequency & 206 & 88 & 23 & 10 & 5 & 3 & 1 
\end{tabular}
\end{table}

The authors propose that the relationship between the frequency,
$N$, of observations and the duration, $D$, in years of a recession
is a power-law
\begin{equation}
N=\frac{\alpha}{D^{\beta}}
\end{equation}
The authors use a non-linear algorithm to determine the best-fit 
parameters $\alpha=209.6$ and $\beta=1.69$. A non-linear 
fit using the Levenberg-Marquardt method reproduces these results, 
yielding parameters $\alpha=209.57$ and $\beta=1.694$ (2 d.p.), with root 
mean square (RMS) error 11.86. The authors note, however, that
an `essential problem of over-prediction of the tail of the
distribution was not solved'.

\begin{figure}[!h]
\centering
\subfigure[The solid line is an exponential fit, $f(D)=\mu \lambda \e^{-\lambda D}$, to the data.]{\epsfig{file=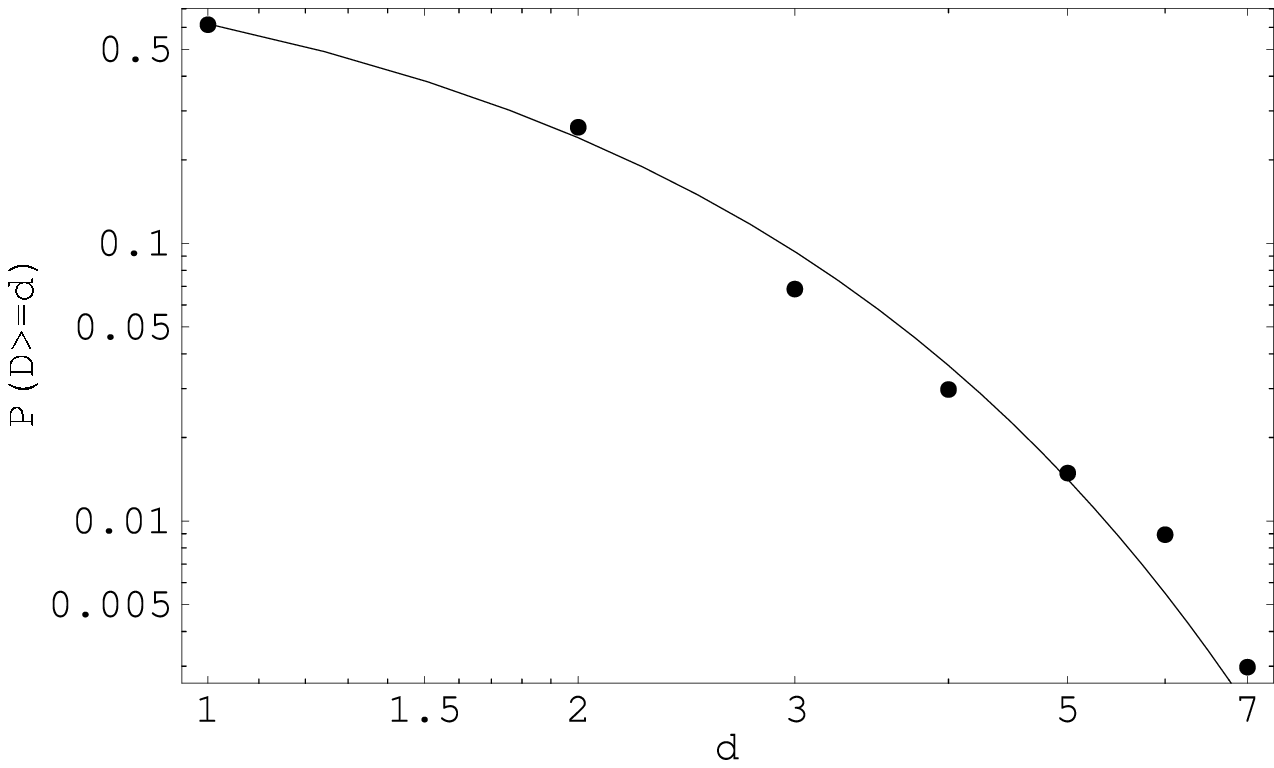,viewport=0 200 634 585,width=0.47\textwidth,clip=true,silent=}}\qquad
\subfigure[The solid line is a power-law fit, $f(D)=\frac{\alpha}{D^{\beta}}$, to the data.]{\epsfig{file=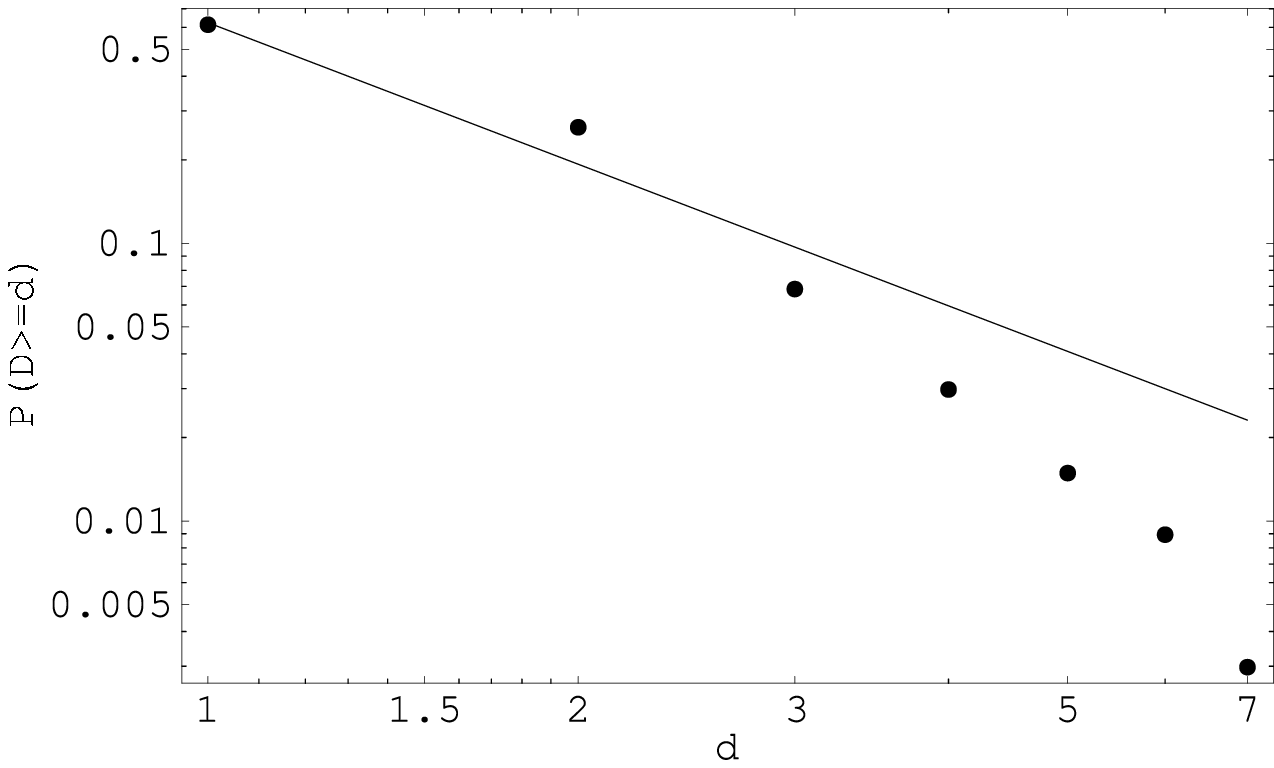,viewport=0 200 634 585,width=0.47\textwidth,clip=true,silent=}}\\
\caption{Graphical analysis of the frequency of duration of recessions plotted as a cumulative probability distribution in log-log scale.}
\end{figure}

The problem of over-prediction of the tail of the distribution
is solved if the relationship between $N$ and $D$ is considered
to be an exponential law
\begin{equation}
N= \mu \lambda \e^{-\lambda D}
\end{equation}
A non-linear fit yields parameters $\mu = 564.85$
and $\lambda = 0.94$, with RMS error 4.67. Figure 1 plots
both the exponential and power-law fits to the data as
complementary cumulative distribution functions (ccdf)
in log-log scale. The data is clearly consistent with
an exponential not a power law.\footnote{The 
frequency of the size of recessions, defined as the cumulative fall 
in output that occured during the recession, is also analysed by the authors.
They conclude that this data follows a power-law.
But in this case also an exponential law yields a lower RMS error than 
a power-law.}

The authors note that if durations of one year are excluded from
the data (representing 61\% of the data) then the power-law fit improves. 
In this case, the exponential fit has a RMS error of 2.01, whereas a power-law
fit has a RMS error 0.57. Figure 2 plots the exponential and power-law
fits to the reduced data. The reduced data is clearly consistent with 
a power-law not an exponential law. 

\begin{figure}[!h]
\centering
\subfigure[The solid line is an exponential fit, $f(D)=\mu \lambda \e^{-\lambda D}$, to the data.]{\epsfig{file=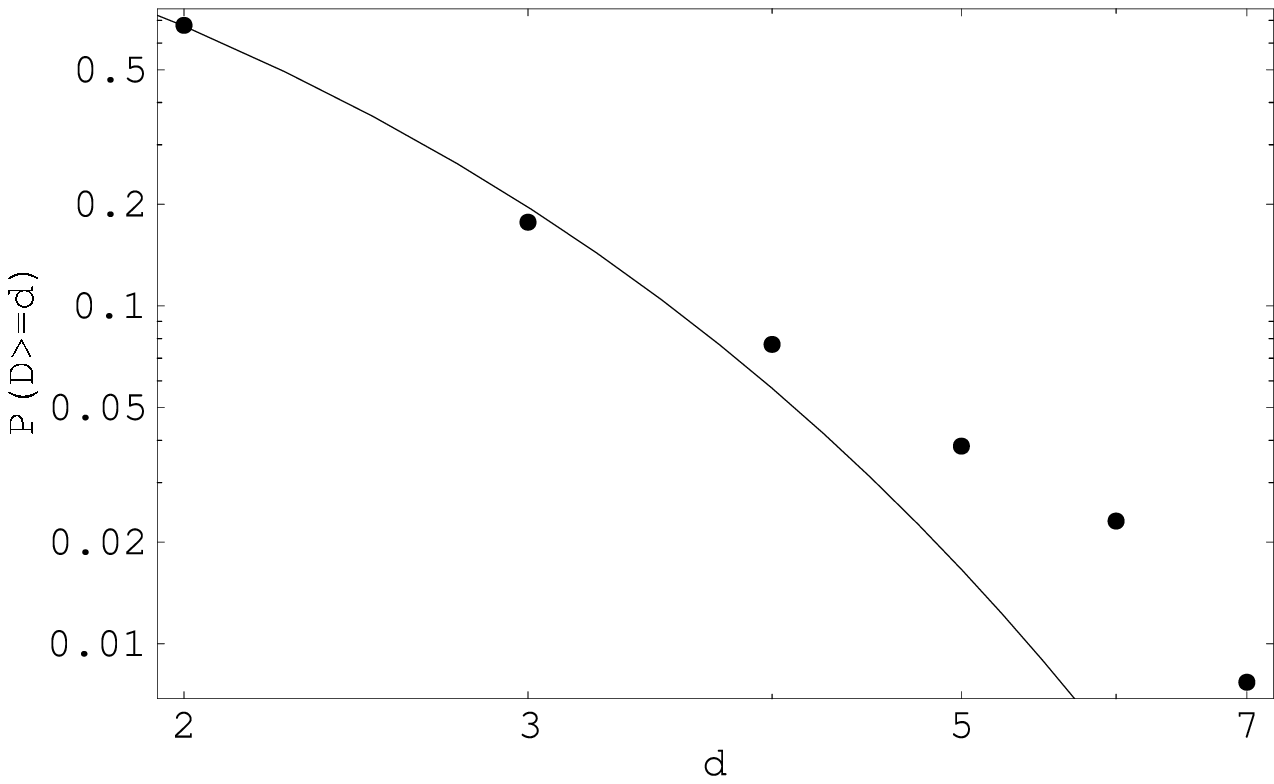,viewport=0 200 634 585,width=0.47\textwidth,clip=true,silent=}}\qquad
\subfigure[The solid line is a power-law fit, $f(D)=\frac{\alpha}{D^{\beta}}$, to the data.]{\epsfig{file=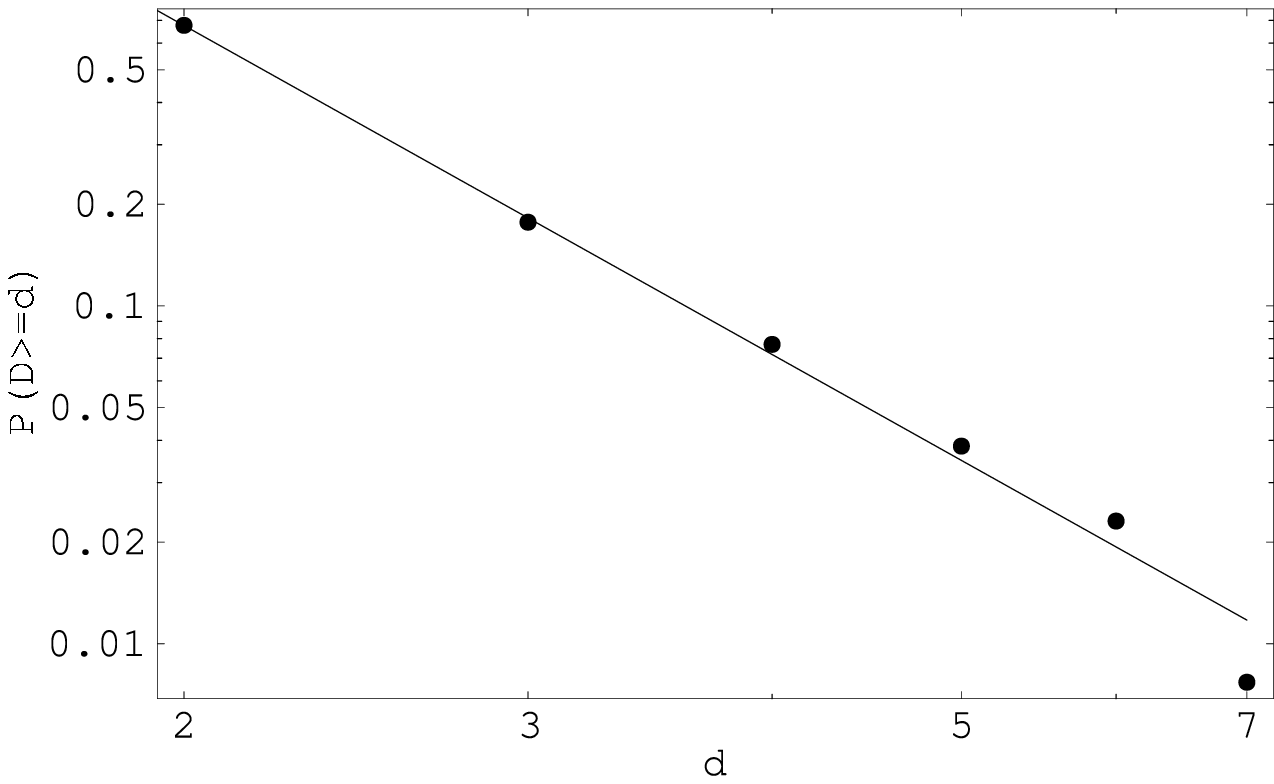,viewport=0 200 634 585,width=0.47\textwidth,clip=true,silent=}}\\
\caption{Graphical analysis of the frequency of duration of recessions that last two years or more plotted as a cumulative probability distribution in log-log scale.}
\end{figure}

The authors propose that economic management often prevents
recessions lasting more than one year, but if they do last longer, then
subjective expectations of growth become depressed and recessions 
may then occur on all scales of duration, resulting in a power-law.
This explanation is indeed consistent with the data once recessions lasting
one year are dropped. But the explanation is redundant once all the data is correctly 
classified as following an exponential distribution.




\end{document}